# Predicted Performance Advantages of Carbon Nanotube Transistors with Doped Nanotubes as Source/Drain

Jing Guo[†], Ali Javey[‡], Hongjie Dai[‡], Supriyo Datta [†] and Mark Lundstrom[†]
[†]School of ECE, Purdue University, West Lafayette, IN, 47907
[‡]Department of Chemistry, Stanford University, Stanford, CA, 94305

*ABSTRACT*

Most carbon nanotube field-effect transistors (CNTFETs) directly attach metal source/drain contacts to an intrinsic nanotube channel. When the gate oxide thickness is reduced, such transistors display strong ambipolar conduction, even when the Schottky barrier for electrons (or for holes) is zero. The resulting leakage current, which increases exponentially with the drain voltage, constrains the potential applications of such devices. In this paper, we use numerical simulations to show that if CNT based metal-oxide-semiconductor (MOS) FETs can be achieved by using heavily doped CNT sections as source and drain, ambipolar conduction will be suppressed, leakage current will be reduced, and the scaling limit imposed by source-drain tunneling will be extended. By eliminating the Schottky barrier between the source and channel, the transistor will be capable of delivering more on-current. The leakage current of such devices will be controlled by the full bandgap of CNTs (instead of half of the bandgap for SB CNTFETs) and band-to-band tunneling. These factors will depend on the diameter of nanotubes and the power supply voltage.

Carbon nanotube field-effect transistors (CNTFETs) with promising device performance have recently been demonstrated [1, 2]. In these transistors, the intrinsic nanotube channel is directly attached to the metal source/drain contacts. Such transistors are referred to as Schottky barrier (SB) CNTFETs and behave like unconventional Schottky barrier transistors [3]. If the metal Fermi level is pinned at the middle of the gap, SB CNTFETs show electron conduction at high gate voltages and hole conduction at low gate voltages. Recently, CNTFETs with zero or slightly negative Schottky barriers were achieved by attaching an intrinsic nanotube channel to the high work function metal contacts [2]. When the gate oxide is thick, reducing the Schottky barrier height to zero suppresses the ambipolar conduction, but when the gate oxide thickness is reduced, the transistor is still ambipolar, even if the Schottky barrier height for electrons/holes is zero [4, 5]. Ambipolar conduction leads to a large leakage current that exponentially increases with the power supply voltage, especially when the tube diameter is large. Very recently, an asymmetric gate oxide SB CNTFET has been proposed as a means of suppressing ambipolar conduction [6]. SB CNTFETs of any type, however, will likely suffer from the need to place the gate electrode close to the source (which increases parasitic capacitance) and metal-induced gap states, which increase source to drain tunneling and limit the minimum channel length.

In this letter, we show that CNT MOSFETs with heavily doped nanotube sections as source/drain will exhibit substantially improved performance. These MOSFET-like CNTFETs will suppress the ambipolar conduction that occurs in SB CNTFETs. They will also extend the channel length scaling limit because of the density of metal-induced-gap-states will be significantly reduced. Under on-state conditions, the MOS CNTFET will operate like a SB CNTFET with a negative Schottky barrier height, which delivers more on-current than SB CNTFETs with positive barrier heights [7]. Finally, the parasitic capacitance between the source and gate electrode will be reduced, which will allow faster operation. MOSFET-like CNTFETs will also display a leakage current in the off-state, but that leakage current is controlled by the full band gap of CNTs and band to band tunneling. The required doping of the source/drain extension may be achievable either chemically [8, 9] or electrically [10]. This letter provides a strong theoretical rationale for developing such devices.

Because our interest is in assessing the ultimate performance capabilities of the two devices, we simulated a coaxially gated SB CNTFET and a MOS CNTFET with a 15nm ballistic channel, as shown in Fig. 1a and 1b, respectively. A 2nm-thick $ZrO_2$ gate oxide was used. (A high-K gate insulator of this type has already been experimentally demonstrated [11].) A power supply voltage of 0.4V was assumed, according to the value specified for the 10nm scale MOSFET in the ITRS roadmap [12]. For the channel, a (13,0) nanotube (diameter, $d \approx 1$ nm, and bandgap $E_g \approx 0.83$ eV) was used.

Both types of CNTFETs were simulated by solving the Schrödinger equation using the non-equilibrium Green's function (NEGF) formalism [13], self-consistently with the Poisson equation. Ballistic transport was assumed. An atomistic description of the nanotube using a tight binding Hamiltonian with an atomistic ($p_z$ orbital) basis was employed. (Note that the computational cost was significantly reduced by using a mode space approach [5, 14].) The charge density was computed by integrating the local density-of-states (LDOS) over energy,

$$Q(z) = (-e)\int_{-\infty}^{+\infty} dE \cdot \text{sgn}[E - E_N(z)]\{D_S(E,z)f\left(\text{sgn}[E - E_N(z)](E - E_{FS})\right)$$
$$+ D_D(E,z)f\left(\text{sgn}[E - E_N(z)](E - E_{FD})\right)\}. \quad (1)$$

Here $e$ is the electron charge, $\text{sgn}(E)$ is the sign function, $E_{FS,D}$ is the source (drain) Fermi level, $E_N(z)$ is the charge neutrality level [15], and $D_{S,D}(E,z)$ is the LDOS due to the source (drain) contact, $D_{S,D} = G\Gamma_{S,D}G^+$, where $G = [(E+i0^+)I - H - \Sigma_S - \Sigma_D]^{-1}$ is the retarded Green's function, $H$ is the device Hamiltonian, $\Sigma_{S,D}$ is the source/drain self-energy, and $\Gamma_{S,D} = i(\Sigma_{S,D} - \Sigma_{S,D}^+)$ is the source/drain broadening function [16].

For SB CNTFETs, the Schottky barriers at the metal/CNT interfaces were treated with a phenomenological source/drain self-energy. To mimic the continuous states injected from metal to the semiconducing nanotube, each semiconducting mode of the channel was coupled to the metallic mode of metallic zigzag CNTs at the M/CNT interface [5]. For MOS CNTFETs, we assumed that the heavily doped source/drain regions were semi-infinite and computed the corresponding self-energy [16].

Along with the NEGF transport equation, we iteratively solve a 2D Poisson equation in cylindrical coordinates. (A non-linear Poisson equation was used to improve the numerical convergence.) Once the self-consistent potential profile was obtained, the source-drain current was computed by

$$I = \frac{4e}{h}\int dE \cdot T(E)[f(E-E_{FS}) - f(E-E_{FD})], \qquad (3)$$

where $T(E) = trace(\Gamma_S G \Gamma_D G^+)$ is the source-drain transmission [16]. The gate leakage current is omitted in this study.

Fig. 2 plots the $I_D$ vs. $V_D$ characteristics for the MOS CNTFET and the SB CNTFETs with different barrier heights. A common off-current of $0.01\mu A$ was specified for all transistors by adjusting the flat band voltage of each transistor individually. As the barrier height of the SB CNTFET decreases, the on-current increases, and finally approaches that of the MOS CNTFET when the barrier height is sufficiently negative. The reason for this behavior was explained in a recent study of silicon SBFETs [7]. For positive SBs, the on-current is limited by the tunneling barrier at the source end of the channel and lowering the barrier height increases on-current. When the barrier height is sufficiently negative, the gate always modulates a thermionic barrier in the bulk nanotube channel, a process similar to the MOS CNTFET, which results in a similar on-current. The advantage of MOS CNTFETs (and negative barrier SB CNTFETs) at on-state is even greater when the gate oxide is thicker, because the thickness of the Schottky barrier is roughly the gate oxide thickness and the thicker Schottky barrier more severely limits the on-current [2].

Next, we explore the off-state performance of CNTFETs in Fig. 2 by sweeping the gate voltage to negative values. Figure 3a plots the $I_D$ vs. $V_G$ characteristics of the MOS CNTFET and the SB CNTFET with a zero height Schottky barrier for electrons. Although the barrier height for holes is high compared to that for electrons, the SB CNTFET still shows strong ambipolar conduction. Fig. 3b, which plots the band profile at the minimal leakage bias, explains the first reason for strong hole conduction. Although the barrier height for holes is the carbon nanotube band gap, the band gap is small compared to that of Si and the tunneling barrier is very thin after the gate oxide thickness is reduced. (An electrostatic calculation shows that the Schottky barrier thickness is roughly the gate oxide thickness.) The second reason for strong hole conduction is that the holes at the valence band edge have strong wave behavior due to the small effective mass. (A parabolic E-k fit of the very bottom of the conduction band of a 1nm diameter CNT gives an effective mass of ~0.08). As a result, the spike-like barrier for holes at the drain end is nearly transparent, and the SB CNTFETs with highly asymmetric barrier heights for electrons and holes still show strong ambipolar conduction when the gate oxide is thin.

The minimal leakage current in Fig. 3a can be estimated by noticing that it occurs when the electron and hole currents are equal. As discussed above, the tunneling barrier for holes at the drain end is nearly transparent when the gate oxide is thin, thus the off-current for holes is limited by thermionic emission over the barrier, $\phi_p$, in the bulk body as shown in Fig. 3b. Equal barrier heights for electrons and holes for electrons and holes are required to produce the same current, therefore, the barrier heights are $\phi_n \sim \phi_p \sim (E_g - eV_D)/2$. By adding the thermionic emission currents for holes and electrons, we find the minimal leakage current as

$$I \sim \frac{8ek_BT}{h} \times \exp(-\frac{E_g - eV_D}{2k_BT}) \qquad (3)$$

in the non-degenerate limit. (Here $h$ is Plank's constant and $T$ is the temperature.) Equation (3) can be interpreted in the following way. At equilibrium, the largest barrier height that limits electron and hole current simultaneous is one half of the band gap, and it decreases by an amount of $eV_D/2$ after the drain voltage is applied.

Figure 3a shows that the minimal leakage current of the MOS CNTFET is several orders of magnitude smaller than the SB CNTFET and that it doesn't strongly depend on the drain voltage. The reason is apparent from Fig. 3c, which plots the band diagram of the MOS CNTFETs at $V_G = -0.3V$. In contrast to the metal contacts, the heavily doped semiconducting source/drain has a band gap energy range for which no states or current are induced into the channel. Because the barrier height for holes in the n$^+$ doped source/drain is roughly the nanotube bandgap, the hole current is negligible and the ambipolar conduction is suppressed. At the same time, the barrier to limit the electron current can approach the nanotube bandgap $E_g$ instead of $E_g/2$ in the SB CNTFET case. (Creating a barrier larger than $E_g$ for electrons may cause band-to-band tunneling as will be discussed later). The minimal leakage current of a MOS CNTFET, therefore, should be greatly reduced from that of a SB CNTFET.

We also compared the scalability of the two transistors and found that the MOS CNTFET was more scalable than the SB CNTFET. Figure 4 shows the $I_D$ vs. $V_G$ characteristics for the SB CNTFET and MOS CNTFET (as shown in Fig. 1) with 5nm channel lengths. Because the metal contacts are directly attached to the intrinsic nanotube channel, a large density of metal-induced-gap-states (MIGS) is produced through the entire 5nm-long channel. Quantum mechanical tunneling from source to drain is severe, and the leakage is large. The on-off current ratio is below 10 and the transistor loses its functionality as a good gate controlled electronic switch. In contrast, for the 5nm MOS CNTFET, the transistor leakage current is substantially smaller, and the on-off ratio is well above 100. This occurs because of the existence of the semiconductor band gap and the corresponding significant reduction in metal induced gap states.

To explore the origin of the leakage current in MOS CNTFETs, we increased the magnitude of the negative gate voltages. Figure 5a, which plots the $I_D$ vs. $V_G$ characteristics for the MOS CNTFETs with three different tube diameters, shows that the drain current increases at high negative gate voltages. Figure 5b, which plots the band profile and the current spectrum schematically, indicates that the large source-drain current at negative gate voltages is due to band-to-band tunneling [8, 17]. When the gate voltage is low, a quantum well is created in the valance band. Electrons in the heavily doped source can tunnel through the eigenstates in the quantum well, which results in the discreet current peaks in the current spectrum. The band-to-band tunneling problem for carbon nanotubes should be more severe than for Si transistors because the band gap is smaller and nanotube is a direct band gap material. The problem is more severe when the tube diameter is larger. As shown in Fig. 5a, the minimal leakage current of the (25,0) CNT is about 5 orders of magnitude larger than that of the (13,0) CNT. (Here we didn't treat single electron charging effects because the thermal energy $k_B T \approx 26 meV$ is larger than the single electron charging energy of $\sim 10 meV$.)

Another advantage of MOS CNTFETs is that the parasitic capacitance between the gate and source/drain electrodes is greatly reduced, which helps the transistor to operate faster. Because the gate modulates the Schottky barrier between the source metal contact and the channel for SB CNTFETs, the gate electrode must be placed close to the source electrode to achieve effective modulation, which, however, results in a large gate/source parasitic capacitance and increases the transistor delay. In contrast, the gate modulates a thermionic emission barrier in the channel region for MOS CNTFETs, the gate and source/drain metal electrodes can be separated by the length of the source/drain extension, which greatly reduces the parasitic capacitance and the transistor delay metric.

We note that the recently developed CNT FETs with high-κ ZrO$_2$ gate insulators and partially gated nanotube in the channel resemble the MOS CNTFETs proposed here. In the experimental case, the un-gated sections are effectively S/D electrodes and heavy p-doping of the sections are unintentional during the ZrO$_2$ deposition process by chloride precursors. More systematic work on the experimental realization of MOS CNTFETs will be presented elsewhere.

In summary, for thin gate oxide devices, the ambipolar conduction of SB CNTFETs cannot be avoided by engineering the Schottky barrier height. Ambipolar conduction results in high leakage currents, especially when the tube diameter is large and the power supply is high. For CNTFETs with heavily doped extensions as source/drain, ambipolar conduction will be suppressed and the leakage will be reduced because the leakage current is limited by thermionic emission over a full band gap rather than a half band gap. At the same time, MOSFET-like CNTFETs will be more scalable than SB CNTFETs. Under on-state conditions, MOS CNTFETs will operate like SB CNTFETs with a sufficiently negative Schottky barrier and will, therefore, deliver more current than normal SB CNTFETs.

## ACKNOWLEDEGEMENT

It is our pleasure to thank Dr Paul Solomon of IBM and Dr. Diego Kienle of Purdue University for helpful discussions. This work was supported by the National Science Foundation, grant no. EEC-0085516, the NSF Network for Computational Nanotechnology, the MARCO Focused Research Center on Materials, Structure, and Devices, and SRC/AMD.

# REFERENCES

[1] S. Wind, J. Appenzeller, R. Martel, V. Derycke, and Ph. Avouris, "Vertical scaling of carbon nanotube field-effect transistors using top gate electrodes," *Appl. Phys. Lett*, vol. 80, pp. 3817-3819, 2002.

[2] A. Javey, J. Guo, Q. Wang, M. Lundstrom and H. Dai, "Ballistic carbon nanotube field-effect transistors," *Nature,* vol. 427, pp. 654-657, 2003.

[3] S. Heinze, J. Tersoff, R. Martel, et al., "Carbon nanotubes as Schottky barrier transistors," *Phys. Rev. Lett.*, vol. 89, pp.106801-04, 2002.

[4] M. Radosavljevic, S. Heinze, J. Tersoff and Ph. Avouris, "Drain voltage scaling in carbon nanotube transistors," cond-mat/0305570, 2003.

[5] Jing Guo, Supriyo Datta and Mark Lundstrom, " A numerical study of scaling issues for Schottky barrier carbon nanotube transistors," cond-mat/0306199, 2003.

[6] S. Heinze, J. Tersoff and Ph. Avouris, "Electrostatic engineering of nanotube transistors for improved performance," cond-mat/0308526, 2003.

[7] Jing Guo and Mark Lundstrom, "A computational study of thin-body, double-gate, Schottky barrier MOSFETs," *IEEE Trans. on Electron Devices*, vol. 49, pp. 1897-1901, 2002.

[8] C. Zhou, J. Kong and H. Dai, "Modulated Chemical Doping of Individual Carbon Nanotubes," *Science*, **290**, 1552 2000.

[9] Jing Kong, Jien Cao, Hongjie Dai, and Erik Anderson, "Chemical profiling of single nanotubes: Intramolecular $p$–$n$–$p$ junctions and on-tube single-electron transistors,." Appl. Phys. Lett., vol. 80, pp. 73-75, 2002.

[10]   S. Wind, J. Appenzeller and Ph. Avouris, "Lateral scaling in carbon nanotube field-effect transistors," cond-mat/0306259, 2003.

[11]   A. Javey, H. Kim, M. Brink, Q. Wang et al. "High-K dielectrics for advanced carbon nanotube transistors and logic," *Nature Materials*, vol. 1, pp. 241-246, 2002.

[12]   International Technology Roadmap for Semiconductors, 2001 Edition, Semiconductor Industry Association, www.itrs.net.

[13]   S. Datta, *Electronic Transport in Mesoscopic Systems* (Cambridge University Press, Cambridge, UK, 1995)

[14]   R. Venugopal, Z. Ren, S. Datta, M. Lundstrom, and D. Jovanovic, "Simulating quantum transport in nanoscale MOSFETs: real vs. mode space approaches," *J. Appl. Phys.*, vol. 92, pp. 3730-3739, 2002.

[15]   J. Tersoff, "Schottky barrier heights and the continuum of gap states," *Phys. Rev. Lett*. vol. 52, pp. 465-568, 1984.

[16]   S. Datta "Nanoscale Device Modeling: the Green's Function Method," *Superlattices and Microstructures,* vol. 28, pp. 253-278, 2000.

[17]   F. Leonard and J. Tersoff, " Multiple functionality in nanotube transistors," *Phys. Rev. Lett*., vol. 88, p. 258302-1, 2002.

# FIGURES

Fig. 1 The simulated, coaxial gate CNTFETs. (a) The SB CNTFET with an intrinsic CNT directly attached to the metal source/drain. (b) The metal-oxide-semiconductor (MOS) CNTFET with the heavily doped source/drain extension. The metal gate electrode is 10nm thick and the source drain doping is $10^9 m^{-1}$ (~0.01 dopant /atom). For both transistors, the $ZrO_2$ gate oxide thickness is 2nm and the dielectric constant 25.

Fig. 2 $I_D$ vs. $V_D$ characteristics at $V_G$ = 0.4V for the MOS CNTFET (the solid line) and the SB CNTFETs (the dashed lines). The off-current of all transistors (defined at $V_D$=0.4V and $V_G$=0) was set at $0.01 \mu A$ by adjusting the flat band voltage for each transistor. For the SB CNTFETs, three barrier heights we simulated: i) at the middle of the gap, $\phi_B = E_g/2$, ii) zero barrier, $\phi_B = 0$, and iii) negative barrier, $\phi_B = -0.3 eV$. The channel is a (13,0) nanotube, which results in a diameter of *d*≈ *1 nm*, and a bandgap of *E$_g$*≈ *0.83 eV*.

Fig. 3 (a) $I_D$ vs. $V_G$ characteristics for the MOS CNTFET (the solid lines) and the zero barrier height ($\phi_{Bn} = 0$) SB CNTFET (the dashed lines) at $V_D$=0.4V and 0.6V. (b) The band profile of the SB CNTFET at the minimal leakage bias ($V_G$=0V) for $V_D$=0.6V. (c) The band profile of the MOS CNTFET when the source-drain current is low. ($V_D$=0.6V and $V_G$=-0.3V). The channel is a (13,0) nanotube.

Fig. 4  $I_D$ vs. $V_G$ characteristics at $V_D = 0.4V$ for the zero barrier $\phi_{Bn} = 0$ SBFET and the MOS CNTFET as shown in Fig. 1. The gated channel of both transistors is a 5nm-long, intrinsic (13, 0) CNT.

Fig. 5 (a) $I_D$ vs. $V_G$ characteristics at $V_D$=0.4V for the MOS CNTFETs with (13,0) CNT channel (the diameter d~1nm), the (17,0) CNT channel (d~1.4nm), and the (25,0) CNT channel (d~2nm). (b) The band profile of the (25,0) CNT (the solid lines) and the schematic plot of the current spectrum at $V_D$=0.4V and $V_G$=-0.6V.

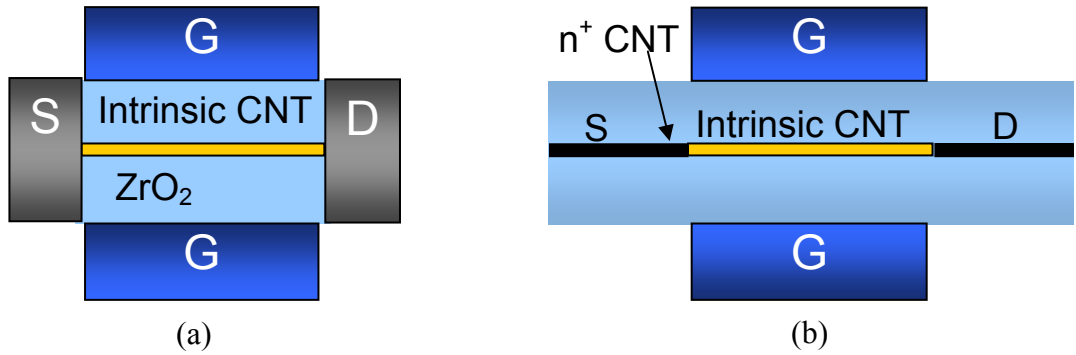

Fig. 1 The simulated, coaxial gate CNTFETs. (a) The SB CNTFET with an intrinsic CNT directly attached to the metal source/drain. (b) The metal-oxide-semiconductor (MOS) CNTFET with the heavily doped source/drain extension. The metal gate electrode is 10nm thick and the source drain doping is $10^9 \, m^{-1}$ (~0.01 dopant /atom). For both transistors, the ZrO$_2$ gate oxide thickness is 2nm and the dielectric constant 25. The channel is a (13,0) nanotube, which results in a diameter of *d*≈ *1 nm*, and a bandgap of *E$_g$*≈ *0.83 eV*.

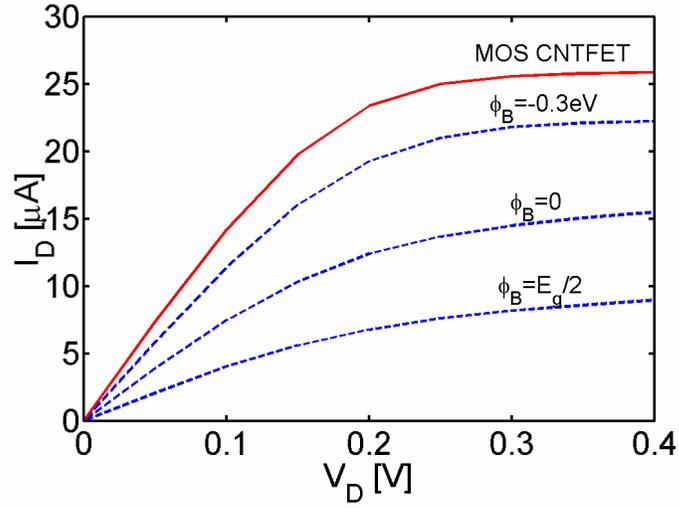

Fig. 2 $I_D$ vs. $V_D$ characteristics at $V_G = 0.4V$ for the MOS CNTFET (the solid line) and the SB CNTFETs (the dashed lines). The off-current of all transistors (defined at $V_D=0.4V$ and $V_G=0$) was set at $0.01\mu A$ by adjusting the flat band voltage for each transistor. For the SB CNTFETs, three barrier heights we simulated: i) at the middle of the gap, $\phi_B = E_g/2$, ii) zero barrier, $\phi_B = 0$, and iii) negative barrier, $\phi_B = -0.3eV$. The channel is a (13,0) nanotube.

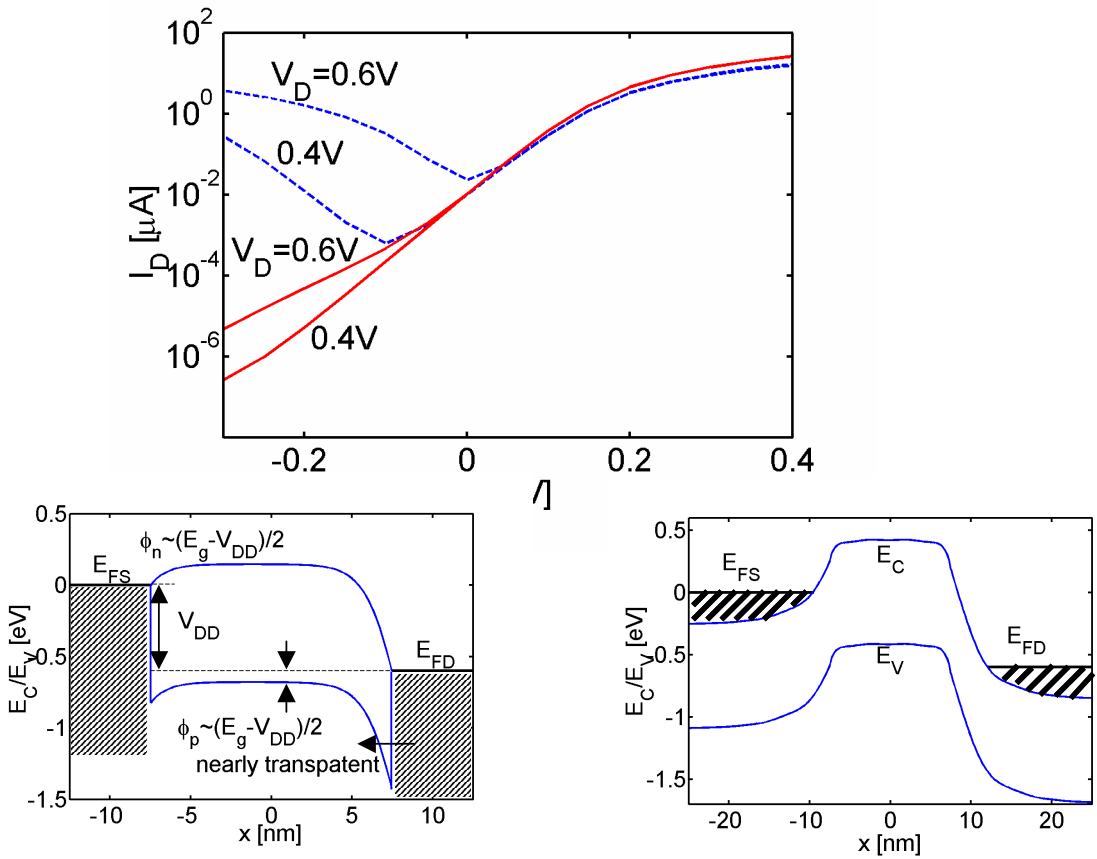

Fig. 3 (a) $I_D$ vs. $V_G$ characteristics for the MOS CNTFET (the solid lines) and the zero barrier height ($\phi_{Bn} = 0$) SB CNTFET (the dashed lines) at $V_D$=0.4V and 0.6V. (b) The band profile of the SB CNTFET at the minimal leakage bias ($V_G$=0V) for $V_D$=0.6V. (c) The band profile of the MOS CNTFET when the source-drain current is low. ($V_D$=0.6V and $V_G$=-0.3V). The channel is a (13,0) nanotube.

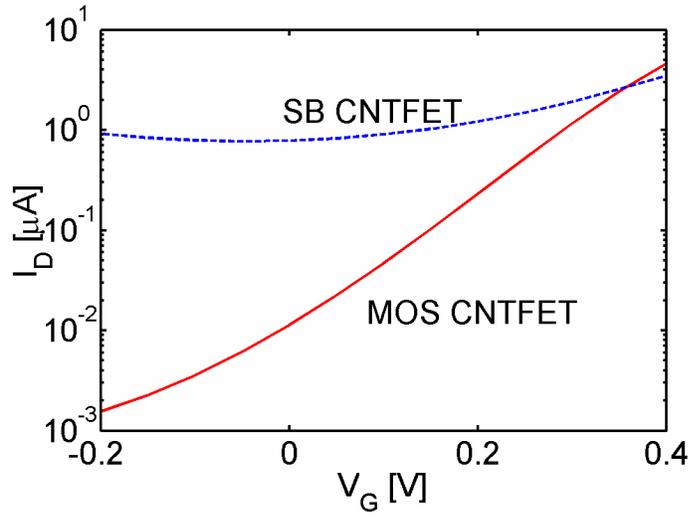

Fig. 4  $I_D$ vs. $V_G$ characteristics at $V_D = 0.4V$ for the zero barrier $\phi_{Bn} = 0$ SBFET and the MOS CNTFET as shown in Fig. 1. The gated channel of both transistors is a 5nm-long, intrinsic (13, 0) CNT.

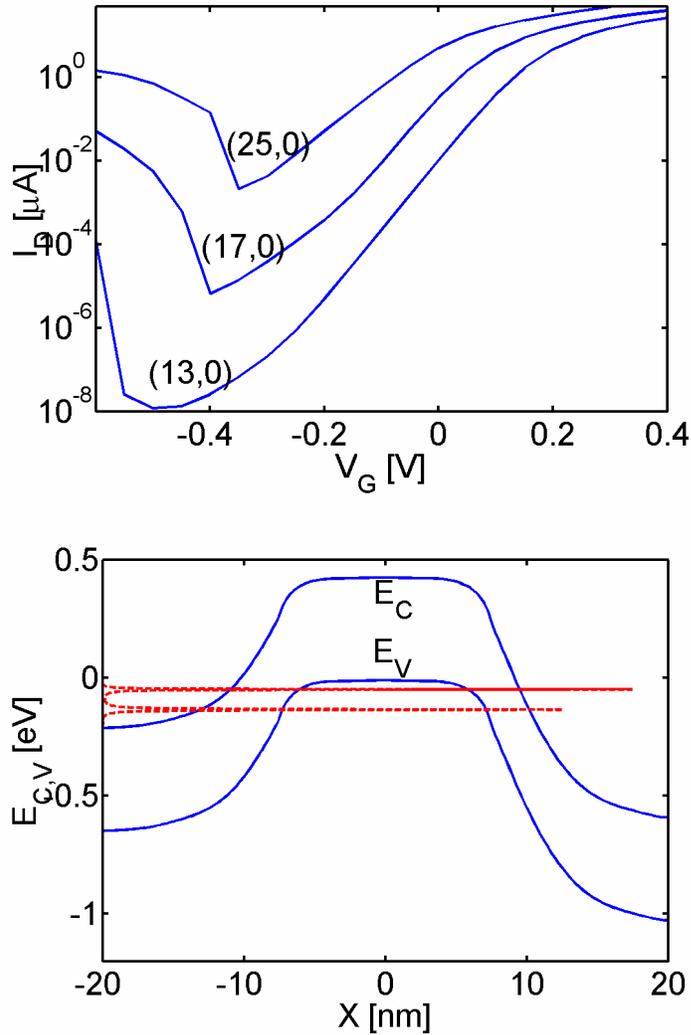

Fig. 5 (a) $I_D$ vs. $V_G$ characteristics at $V_D$=0.4V for the MOS CNTFETs with (13,0) CNT channel (the diameter d~1nm), the (17,0) CNT channel (d~1.4nm), and the (25,0) CNT channel (d~2nm). (b) The band profile of the (25,0) CNT (the solid lines) and the schematic plot of the current spectrum at $V_D$=0.4V and $V_G$=-0.6V.